\documentclass[aip, preprint]{revtex4-1}

\usepackage{hyphenat}
\usepackage{amsmath}
\usepackage{comment}
\usepackage{float}
\usepackage{graphicx}
\usepackage{textcomp}
\usepackage[version=4]{mhchem}
\usepackage{gensymb}
\usepackage{hyperref, color}
\definecolor{linkColor}{rgb}{0.8,0,0}
\definecolor{darkred}{rgb}{0.8,0,0}
\hypersetup{pdfborder={0 0 0},colorlinks=true,urlcolor=linkColor,citecolor=linkColor}

\usepackage{fancyhdr}
\begin{document}

\title{Mind the Gap: Bridging the Divide Between AI Aspirations and the Reality of Autonomous Characterization}

\author{Grace Guinan}
\affiliation{National Renewable Energy Laboratory, Golden, Colorado 80401}

\author{Addison Salvador}
\affiliation{National Renewable Energy Laboratory, Golden, Colorado 80401}
\affiliation{University of Cincinnati, Cincinnati, Ohio 45221}

\author{Michelle A. Smeaton}
\affiliation{National Renewable Energy Laboratory, Golden, Colorado 80401}

\author{Andrew Glaws}
\affiliation{National Renewable Energy Laboratory, Golden, Colorado 80401}

\author{Hilary Egan}
\affiliation{National Renewable Energy Laboratory, Golden, Colorado 80401}

\author{Brian C. Wyatt}
\affiliation{School of Materials Engineering, Purdue University, West Lafayette, Indiana 47907}

\author{Babak Anasori}
\affiliation{School of Materials Engineering, Purdue University, West Lafayette, Indiana 47907}
\affiliation{School of Mechanical Engineering, Purdue University, West Lafayette, Indiana 47907}

\author{Kevin R. Fiedler}
\affiliation{National Security Directorate, Pacific Northwest National Laboratory, Richland, Washington 99352}

\author{Matthew J. Olszta}
\affiliation{Energy and Environment Directorate, Pacific Northwest National Laboratory, Richland, Washington 99352}

\author{Steven R. Spurgeon}
\email{steven.spurgeon@nrel.gov}
\affiliation{National Renewable Energy Laboratory, Golden, Colorado 80401}
\affiliation{Metallurgical and Materials Engineering Department, Colorado School of Mines, Golden, Colorado 80401}
\affiliation{Renewable and Sustainable Energy Institute, University of Colorado Boulder, Boulder, Colorado 80309}

\date{\today}


\begin{abstract}

What does materials science look like in the ``Age of Artificial Intelligence?'' Each materials domain---synthesis, characterization, and modeling---has a different answer to this question, motivated by unique challenges and constraints. This work focuses on the tremendous potential of autonomous characterization within electron microscopy. We present our recent advancements in developing domain-aware, multimodal models for microscopy analysis capable of describing complex atomic systems. We then address the critical gap between the theoretical promise of autonomous microscopy and its current practical limitations, showcasing recent successes while highlighting the necessary developments to achieve robust, real-world autonomy.

\end{abstract}

\maketitle

\section{Introduction}

Our society has rapidly entered the ``Age of Artificial Intelligence (AI),'' with new AI-infused products and research emerging weekly, many of uncertain value. This trend is mirrored in the materials science community,\cite{Choudhary.10.1038/s41524-022-00734-62w, Sha.10.1002/aisy.201900143,Butler.10.1038/s41586-018-0337-2,Batra.10.1038/s41578-020-00255-y} which has embraced these approaches while simultaneously striving to assess their true worth. Many of the most noteworthy successes are found in accelerated prediction of novel materials \cite{merchant2023scaling,chen2024accelerating} aided by modeling, often in cases where machine learning (ML)-based potentials are accelerating computational materials science by enabling simulations at previously inaccessible scales.\cite{Kadupitiya.10.1016/j.jocs.2020.101107,Fedik.10.1038/s41570-022-00416-3}. Similarly compelling successes are found in materials synthesis,\cite{Pyzer-Knapp.10.1038/s41524-022-00765-z,Harris.10.1002/smtd.202301763} which is being transformed by fast, autonomous decision-making. Between these fields lies materials characterization, which acts as the arbiter between the tidy, idealized world of modeling and the messy reality of synthesis.\cite{Kalinin.10.1038/s41524-023-01142-0,Kaufmann.10.1016/j.cossms.2024.101192} By effectively implementing AI in characterization, we gain the opportunity to conduct more informative, reproducible, and effective experiments, informing greater control of materials synthesis and processing.

The diverse sub-fields of materials characterization present unique challenges and opportunities for AI/ML. Scanning transmission electron microscopy (STEM), in particular, is a cornerstone of materials science that stands to be profoundly impacted by AI/ML for several reasons.\cite{Spurgeon.10.1038/s41563-020-00833-z,Kalinin.10.1557/s43577-022-00413-3, Treder.10.1093/jmicro/dfab043} First, electrons are exceptionally rich probes of materials due to their strong interaction with matter. STEM experiments generate imaging, spectroscopic, and diffraction signals that collectively form a multimodal, time-varying descriptor across millimeters to picometers.\cite{Smeaton.10.1021/acsnano.4c09256} Here the multimodal aspect of the data arises from the varying methods for probing electron interaction mechanisms, leading to data spanning dimensionality from spectra to images to 4D-STEM data. Second, STEM techniques have historically been labor-intensive and heavily reliant on human expertise.\cite{Bruno.10.1093/micmic/ozad067.767, Kalinin.10.1093/mictod/qaad096} This reliance has hindered our ability to reproduce and scale experiments, owing to both the inherent ``art'' of microscope operation and the often opaque nature of instrument control. Finally, recent hardware advancements have produced ultrafast cameras and detectors capable of generating immense, multimodal datasets that far exceed the limits of human cognition.\cite{Spurgeon.10.1038/s41563-020-00833-z} We must now determine how best to apply emerging data science approaches and identify areas where they can have the greatest impact.

A longstanding goal of materials science has been to develop ontologies for understanding matter across vast spatial, chemical, and temporal scales---in essence, we aim to create robust \textit{descriptive} statistics.\cite{Norouzi.10.48550/arxiv.2408.06034,Buehler.10.1088/2632-2153/ad7228} STEM allows us to probe the constituents of matter with high fidelity, informing these ontologies, but we must carefully consider the nuances and interpretations associated with each data modality. For instance, the apparent width of a grain boundary or interface may differ depending on whether it is measured using imaging or spectroscopic techniques. Furthermore, the physics of electron-sample interactions leads to signal delocalization and complexities in data interpretation.\cite{Oxley.10.1103/physrevb.76.064303, Spurgeon.10.1017/s1431927617000368} Therefore, to inform such an ontology using microscopy, we require approaches that: 1) can operate in sparse discovery scenarios, where \textit{a priori} information about features of interest may be limited; 2) are amenable to physical interpretation; and 3) can harness multimodal data that collectively reveal different facets of structure, chemistry, and composition. Fortunately, recent advancements in ML have begun to address these challenges. For example, high-throughput physics-based forward modeling enables the construction of synthetic data for model training in well-bounded scenarios.\cite{Dan.10.1126/sciadv.adj0904,Lin.10.1038/s41598-021-84499-w, Schwenker.10.1002/smll.202102960} Alternatively, few-shot classification approaches can rapidly triage data in otherwise intractable situations.\cite{Akers.10.1038/s41524-021-00652-z, Fujinuma.10.1038/s43246-022-00283-x, Kaufmann.10.1038/s41598-021-87557-5, Chen.10.1038/s41598-023-29606-9} Computer vision techniques, aided by packages like AtomAI\cite{Ziatdinov.10.1038/s42256-022-00555-8} and MicroNet,\cite{Stuckner.10.1038/s41524-022-00878-5} have significantly improved the accuracy and scalability of quantifying atomic structure, thereby informing physical models of point defects. Finally, emerging multimodal models\cite{Schwartz.10.1038/s41524-021-00692-5, Ter-Petrosyan.10.48550/arxiv.2411.09896, Ning.10.1016/j.matt.2024.04.030} are integrating the full spectrum of data generated by modern instruments to create truly representative descriptors of crystalline order.

Ultimately, as materials scientists, our goal extends beyond mere \textit{description}; we aim to \textit{prescribe} material responses. In essence, we aspire to use the microscope for both measurement and manipulation. Achieving this requires transitioning from post-hoc data analysis to real-time classification, control, and feedback within our instrumentation. Over the past decade, significant progress has been made in the engineering challenge of shifting from purely human-in-the-loop to autonomous, closed-loop STEM experiments.\cite{Roccapriore.10.1021/acsnano.2c07451,Dyck.10.1038/s41578-019-0118-z,Dyck.10.1002/smll.201801771} These experiments have begun to enable statistical studies at unprecedented scales, characterizing the behavior of millions of atoms and defects and validating synthesis across thousands of particles. Autonomous control of the electron beam, stage, and environment allows us to program materials atom-by-atom, precisely positioning dopants and introducing vacancies,\cite{Trentino.10.1021/acs.nanolett.1c01214,Dyck.10.1063/1.4998599, Lewis.10.1038/s41524-022-00940-2} all guided by real-time ML models and closed-loop control.

\begin{figure}[h!]
    \centering
    \includegraphics[width=\textwidth]{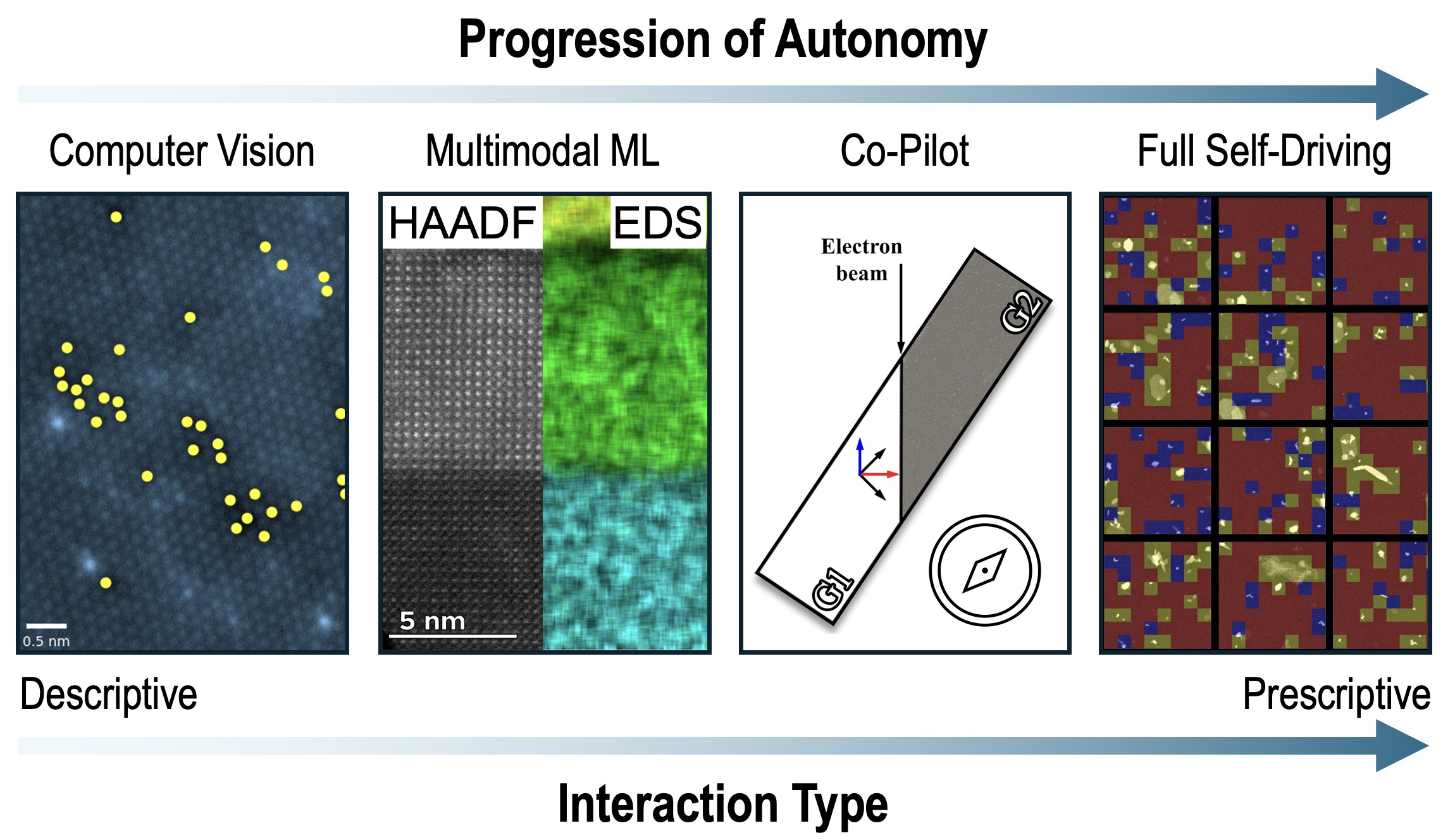}
    \caption{\textbf{The quest for the autonomous microscope.} Bridging from descriptive to prescriptive analytics is a central challenge in achieving full autonomy. Second panel adapted from Ter-Petrosyan et al.\cite{Ter-Petrosyan.10.48550/arxiv.2411.09896}. Third panel adapted from Olszta et al.\cite{Olszta.10.69761/dnka1581} under CC-BY 4.0 license.}
    \label{fig:overview}
\end{figure}

Despite this progress, numerous roadblocks to practical autonomy persist, ranging from models unsuitable for discovery scenarios to the frustrating limitations in programming our own instruments.\cite{Kalinin.10.1038/s41524-023-01142-0,Kalinin.10.1557/s43577-022-00413-3} Our group has dedicated the better part of the last decade to developing autonomous microscopy, informed by sparse and multimodal approaches, time series forecasting, and closed-loop control. In this perspective, we aim to introduce new scientists to the field, focusing on four primary topics spanning descriptive to predictive autonomy, as illustrated in Fig. \ref{fig:overview}. These include the development of domain-specific computer vision, multimodal models capable of harnessing a wide array of signals, co-pilots to aid in instrument control, and fully autonomous systems guided by ML agents. While many studies omit implementation details, we place particular emphasis on explaining the rationale behind our approaches and the real-world challenges of autonomous experimentation. Our intention is not to discourage new practitioners, but rather to demystify the art of autonomy, highlight early successes, and identify key areas ripe for future development. We invite the materials science community to join us in embracing the challenges and opportunities of AI-driven characterization.

\section{Building the Materials Ontology with Autonomous Microscopy}

The development of materials ontologies---systematic mappings of data to concepts and relationships---lies at the heart of materials science.\cite{Baas.10.1109/ACCESS.2023.3327725} Whether forging steel, depositing microelectronic interconnects, or spin-coating polymers, our aim is to relate processing to observations of real-world structure to gain mastery over matter. While the creation of such ontologies is a community-wide endeavor, for brevity and focus, we restrict our discussion to STEM, which nevertheless provides information-dense, multi-scale, and multimodal probes of materials.

A natural question arises: why prioritize ontology development? For millennia, humans have steadily improved materials through Edisonian, trial-and-error methods. However, the past two decades have witnessed an explosion in our capacity to synthesize, characterize, and model materials. A century ago, a single researcher might have manually analyzed a handful of optical images; today, we can easily generate thousands of images, spectra, and diffraction patterns in a single hour---a dataset that would require more than a human lifetime to fully analyze. This presents a critical decision point: either we concentrate our manual efforts on a small subset of (hopefully) representative data, or we leverage the capacity of ML models to analyze large, collective data in a tireless, reproducible manner. By choosing the latter, we can begin to bridge spatial, chemical, and temporal scales, informing a truly data-driven ontology with enhanced descriptive and prescriptive power.

Here, we present two examples of how ML can significantly contribute to this ontological development. First, we consider the synthesis of 2D materials, aiming to understand and map the spatial distribution of point defects statistically using computer vision. Second, we examine order in irradiated crystalline oxides through multimodal imaging and spectroscopy. In both cases, our goal is to bridge the gap between high-resolution insights, statistically relevant datasets, and experimental reproducibility.

\subsection{Atomically Precise Defect Discovery}\label{sec:atom-finding}

Materials science is as much focused on controlling the presence of atoms as it is controlling their absence. Many nanostructured materials, such as thin films and 2D materials, are strongly defined by the presence of point defects such as interstitials and vacancy atoms.\cite{Schaibley.10.1038/natrevmats.2016.55,Novoselov.10.1126/science.aac9439, Gunkel.10.1063/1.5143309} The latter defect is particularly challenging to detect, since vacancies usually occur in a dilute, aperiodic fashion (unless ordering in superstructures\cite{Wang.10.1063/1.5096769x0d}) and induce only subtle lattice distortions in the surrounding crystal. As a result, STEM approaches are one of the few techniques able to capture these defects. Here human-in-the-loop experimentation poses a challenge, as we are quite often looking for needles in a very noisy haystack. To date, some of the most successful implementations of ML have been in the study of 2D graphene materials, where computer vision has been used to detect and quantify impurity atoms and vacancies with great statistical rigor.\cite{Ziatdinov.10.1126/sciadv.aaw8989} However, these approaches have not yet been fully embraced in other 2D materials and to capture the topology of more sophisticated defect ensembles.

One such class of materials where ML can make a particularly strong impact is MXenes, 2D inorganic compounds with the general formula of $\mathrm{M}_{n+1}\mathrm{X}_n\mathrm{T}_x$, where $M$ is an early transition metal, $X$ is carbon and/or nitrogen, and $T$ is a surface functional group (e.g., --O, --OH and --F).\cite{Lukatskaya.10.1038/ncomms12647,Wyatt.10.1038/s41578-023-00619-0} Since their discovery in 2011,\cite{Naguib.10.1002/adma.201102306} the MXenes have attracted considerable attention for their unique properties, with the usage of electrochemical etching as a possible means to control their point defect content.\cite{Sang.10.1021/acsnano.6b05240} However, to date, we have been limited to describing defect distributions using manual analysis of a handful of representative images. We now have the opportunity to apply ML approaches to perform these analyses, potentially 1) increasing the quantity of analyzed images far beyond what was than previously feasible and 2) increasing the speed, precision, and reproducibility of these analyses.

\begin{figure}[h]
    \centering
    \includegraphics[width=\textwidth]{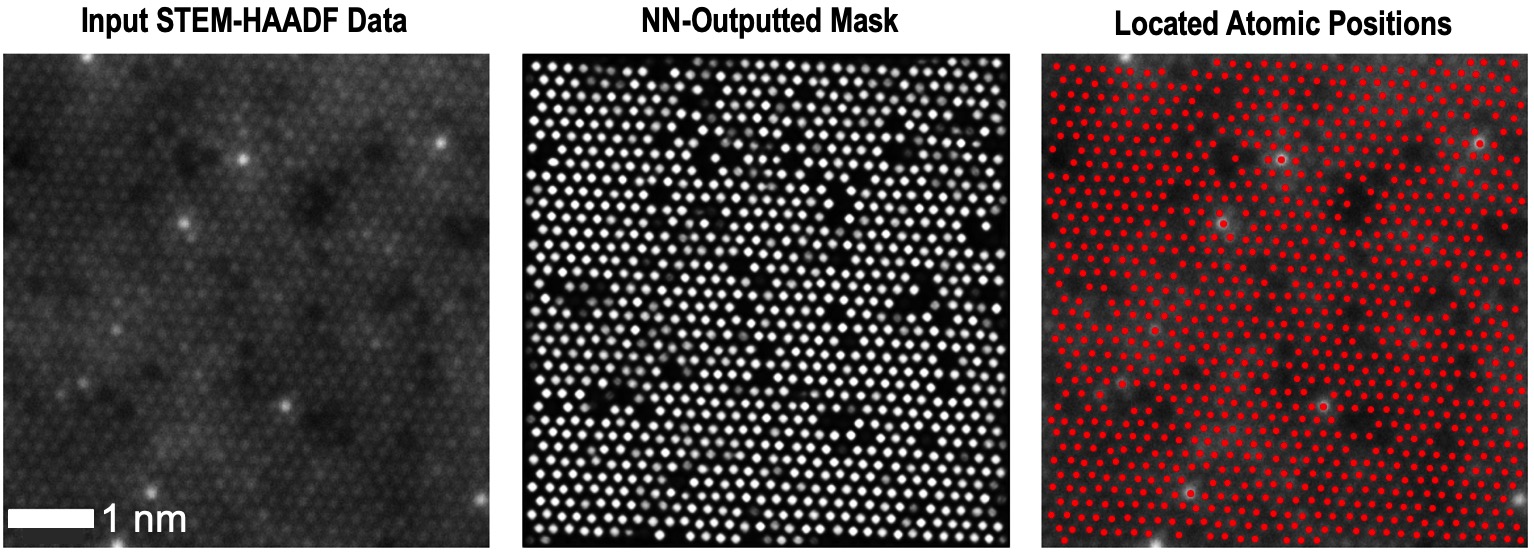}
    \caption{\textbf{Computer vision reveals atomic positions in 2D materials.} Raw STEM-HAADF MXene images (left) are translated into atomic positions using a Neural Network (NN). The NN first outputs a mask (center) and then finds atomic positions (right). Through ML we are able to analyze MXene images in a more reproducible, scalable manner.}
    \label{fig:precision}
\end{figure}

Here, we introduce an ongoing study that uses ML to locate and classify point defects in MXenes, as shown in Fig. \ref{fig:precision}, with the goal to describe the defect topology. As this research is ongoing, we will discuss our general ML thoughts and strategies that are applicable across the board when looking to apply ML techniques to microscopy data. We focus on three aspects of ML: creating labeled training datasets, constructing model-specific data collection strategies, and evaluating model performance.

Our first challenge is a fundamental ML question: how can we generate labeled training data for our model? To locate atomic positions in experimental STEM images, we used the aforementioned AtomAI, a framework for deep learning in microscopy.\cite{Ziatdinov.10.1038/s42256-022-00555-8} AtomAI has some pretrained segmentation models, but these were unsuccessful on our MXene data as they were trained on different classes of materials. Therefore, we turned to training our own model. Ideally, we would train on a large labeled experimental dataset; however, these are currently lacking in materials science and nonexistent for MXenes. We solved this issue by sourcing a previously published wide field of view image,\cite{Sang.10.1021/acsnano.6b05240} randomly cropping the image, and locating atomic positions using Gaussian fitting. Since we trained on a single large image, it was difficult to generalize the model broadly. Instead, we chose to optimize our model's parameters to succeed within a range of resolutions, magnifications and other parameters by altering our training data to these specifications and adding data augmentation (specifically, rotation and Gaussian noise). Simulated data are another way to solve this problem, as efficient STEM multislice simulations\cite{Pelz.10.1017/s1431927621012083} can rapidly produce a large amount of training data with varying parameters (e.g., lattice orientation, resolution, magnification, tilt, etc.).\cite{Eliasson.10.1021/acs.nanolett.4c06025,DaCosta.10.1038/s41524-024-01336-0,Leitherer.10.1038/s41524-023-01133-1,Khan.10.1038/s41524-023-01042-3} However, this can also lead to inaccuracies if the synthetic data does not well approximate the experiment.

Next, we focus on co-design of data collection with the ML architecture in mind. Uniformly collected and stored data and metadata is key in ML; however, currently data are not captured with ML in mind. Data scientists and materials scientists must work together to establish a standardized framework of data and metadata storage for models to pull from. It is also important to understand the model's limitations to collect data it can efficiently process. Models often have specific requirements for consistency in image resolution, size (commonly equal to $2^n$), and magnification. To address this, we created a resolution and magnification series to test our model's limitations, where we focused on one region and independently increased each parameter. These series demonstrate where the neural network works and where it breaks; through this, we can derive concrete requirements for imaging resolution, magnification, detector settings, etc. To capture useful data for ML, it is important to understand limitations of the model and set guidelines for advantageous image acquisition and metadata storage.

Finally, we have found it is difficult to validate model performance because of the lack of universal ground truths. This finding reinforcesour first point about labeled training data. Traditionally, data scientists sequester a portion of their training data for post-training model validation, but this method is impractical for materials scientists given the lack of training data. Past work compares model outputs to an expert's hand-labeling, which is known to be problematic, as well as tedious and time-consuming.\cite{Bruno.10.1093/micmic/ozad067.767} For these reasons, model validation is still an open question in materials science. 

Currently, ML is not as widely used in materials science compared to other fields, partially due to the lack of time and money to develop these methods. 
However, ML in materials science will benefit from increased automation. Automation and ML can work hand-in-hand, with ML feeding relevant information to the automated system and automation providing large quantities of data to feed back into ML models. For example, it is currently difficult to efficiently obtain a massive amount of images in a uniform manner; emerging automated image acquisition will open the door to constructing large datasets that are tailored to specific ML models. Since they are so interconnected, it is important that we develop these methods in parallel.

ML methods are often presented as magical solutions, but they are not always a matter of straightforward application. Still, these methods present an exciting opportunity to advance materials science, through their unique ability to increase the scale and speed of analyses. Our MXene research is only made possible through ML, which allows us to calculate true representative statistics through fast defect detection and categorization across many images.

\subsection{Addressing the Inverse Problem with Multimodal Microscopy}\label{sec:multimodal}

A fundamental goal of materials science is to solve the inverse problem: translating downstream measurements into an atomic or chemical structure, ideally with an associated confidence metric. In the context of electron microscopy, we can consider a simplified linear image approximation: $g(x,y) = f(x,y) \otimes h(x,y)$.\cite{Bertero.10.1201/9781003032755} Here, $f(x,y)$ represents the object function—the ideal image of the object (e.g., a crystal) we wish to determine. $h(x,y)$ is the point spread function, which accounts for aberrations introduced by the imaging system. Finally, $g(x,y)$ represents our recorded data. Unfortunately, aberrations, noise, and other imaging artifacts prevent a one-to-one mapping between the image and the object, leading to information loss. Consequently, we can only obtain approximations of the object function, introducing inaccuracies in our solutions to the inverse problem. As previously mentioned, electron microscopes exploit the strong interactions between the incident electron beam and the sample, generating a variety of elastically and inelastically scattered signals.\cite{Spurgeon.10.48550/arxiv.2001.00947} These signals are each affected by different imaging artifacts, provide information about different sample characteristics, and can be modeled using distinct physics-based approaches. By leveraging this multimodal data, we may achieve more accurate and confident solutions to the inverse problem.

The collective analysis of multimodal STEM data has long been challenging for three primary reasons. First, traditional human-in-the-loop experiments have relied on manual, often iterative, inspection of data. While this approach has yielded valuable correlations, such as the relationship between structural distortions observed in imaging and chemical state changes revealed by spectroscopy, it is time-consuming and can overlook subtle, latent correlations that are not immediately obvious. Second, while we often assume that more data is better, this overlooks the cost associated with acquiring STEM data. This cost can manifest as acquisition time, but more critically, it can involve sample alterations induced by the imaging process itself, particularly when data is acquired sequentially. For example, imaging doses might be relatively low ($\sim10^3~e~\text{\r{A}}^{-2} $), whereas electron energy loss spectroscopy (EELS) doses can be significantly higher ($\sim 10^5~e~\text{\r{A}}^{-2}$).\cite{Smeaton.10.1021/acsnano.4c09256} Third, applying ML approaches to this problem reveals the need for novel architectures and encoders capable of handling multimodal, multi-fidelity, and time-varying data. Only recently have robust electron microscopy image-specific encoders been developed, and we still lack widely accepted encoders for spectroscopic and diffraction data. Furthermore, the optimal structure and classification methods for such data remain unclear, as does the extraction of physically meaningful descriptors.

To address these challenges, we investigated the evolution of order in complex oxide interfaces subjected to heavy-ion irradiation. These materials are widely used in power electronics, sensors, and computing architectures, and their properties are intrinsically linked to local crystalline and chemical order.\cite{Matthews.10.1021/acs.nanolett.1c01651,Spurgeon.10.1002/admi.201901944,Spurgeon.10.1016/j.cossms.2020.100870, Prameela.10.1038/s41578-022-00496-z} In our study, we tracked the evolution of STEM high-angle annular dark-field (STEM-HAADF) images, which reflect both the underlying lattice structure and provide semi-quantitative compositional information.\cite{Ter-Petrosyan.10.48550/arxiv.2411.09896} Concurrently, we performed STEM energy-dispersive X-ray spectroscopy (STEM-EDS) analysis to map local compositional changes resulting from oxygen loss and ballistic mixing. Our objective was twofold: to extract a quantitative, statistical descriptor of disorder and to correlate it with the structural and chemical properties of the material.

\begin{figure}[H]
    \centering
    \includegraphics[width=0.66\textwidth]{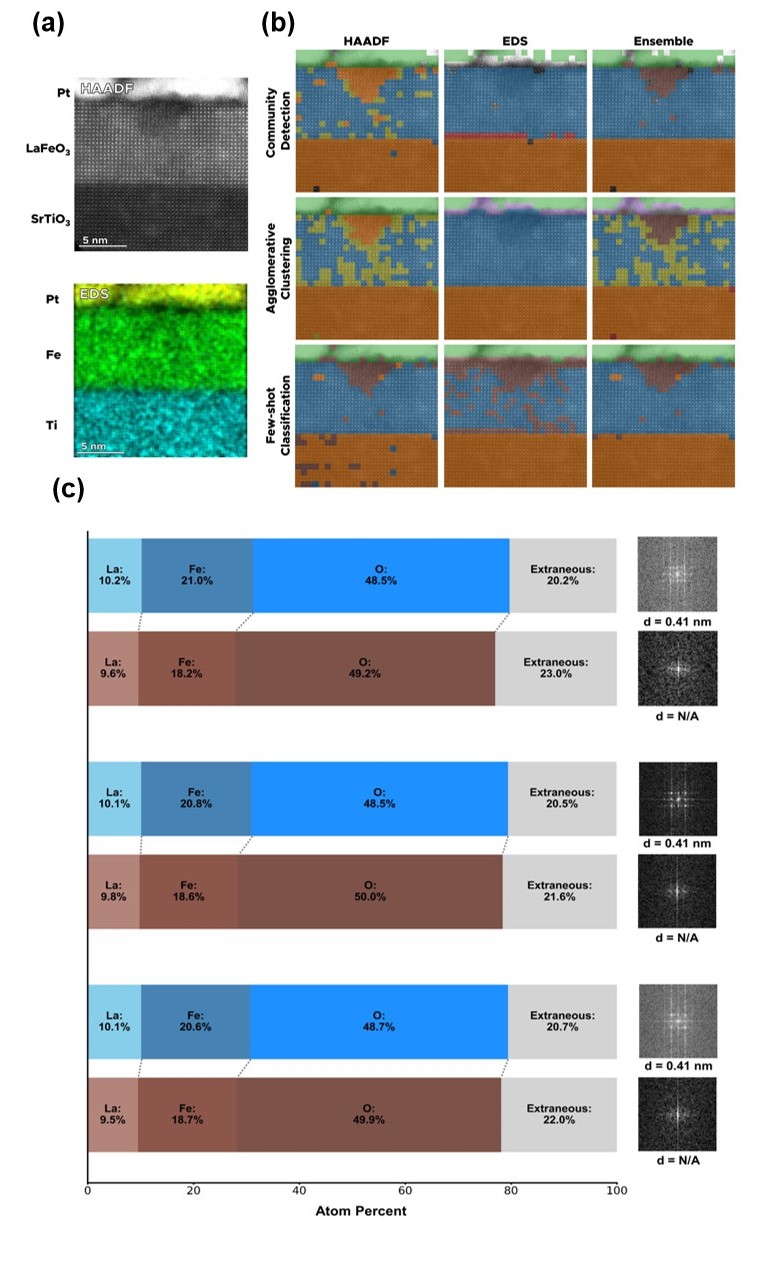}
    \caption{\textbf{Multimodal analytics informs material descriptors.} (a) STEM-HAADF and STEM-EDS provide structural and compositional information on disorder in oxides. (b) Various data classification approaches show the improved discriminating power of multimodal data. (c) These classifiers inform physical descriptors for disorder, including compositional changes and crystallinity. Adapted from Ter-Petrosyan et al.\cite{Ter-Petrosyan.10.48550/arxiv.2411.09896}}
    \label{fig:multimodal}
\end{figure}

As illustrated in Fig. \ref{fig:multimodal}.(a), imaging and spectroscopic data provide complementary information about the sample. STEM-HAADF imaging is lattice-resolved and readily reveals amorphous regions within the crystal, as well as subtle changes in the surrounding lattice and interfaces. STEM-EDS spectroscopy, while having somewhat lower spatial resolution, offers direct quantification of the elemental distribution. Fig. \ref{fig:multimodal}.(b) demonstrates the performance of a classifier incorporating both unimodal and multimodal data for three different classification approaches, with increasing levels of manual fine-tuning from top to bottom. Our key finding is that neither imaging nor spectroscopy alone adequately describes the disordered region in the center of the crystal. While imaging can delineate the boundaries around the disordered region, it is susceptible to noise arising from contrast variations induced by sample preparation. Conversely, STEM-EDS exhibits greater sensitivity to strong compositional boundaries (e.g., segregation and layering) and less sensitivity to contrast variations around the disordered region. An ensemble model effectively discriminates regions of disorder, with performance varying depending on the chosen classification approach. From these resulting segmentation masks, we can extract physical properties of the crystal, as shown in Fig. \ref{fig:multimodal}. Specifically, we find that the degree of structural disorder can be quantified through fast Fourier transforms (FFTs) of the crystal lattice, a standard technique for assessing lattice parameters. Simultaneously, we observe changes in oxygen content within the disordered regions, resulting from the formation of irradiation-induced oxygen vacancies. Collectively, this approach yields a potentially more statistical and reproducible descriptor of disorder.

While this study represents an important step forward in using multimodal ML to characterize crystalline order, we acknowledge several limitations. First, sophisticated and standardized encoders for microscopy data, particularly STEM-EDS spectroscopy, are lacking. Such encoders are crucial for accurately representing features and can significantly influence the outcome of any classification task. Second, we observed performance differences depending on how the data were ensembled, suggesting a trend toward modality bias. This highlights the importance of using balanced datasets, which is challenging given the relative sparsity of STEM-EDS data compared to imaging data. Regularization strategies like dropout, engineered loss functions that utilize all available modalities, and learned fusion mechanisms can be employed to address these issues. This represents a crucial area for future development in multimodal models. Nevertheless, these preliminary results demonstrate the power of ML approaches to overcome the limitations of human-in-the-loop experimentation, offering a scalable, reproducible metric for describing order in crystalline materials based on STEM data.

\section{Navigating the Nanoworld: Practical Autonomy in Electron Microscopy}

The preceding sections have highlighted the power of ML in interrogating the atomic world and extracting meaningful descriptors from high-resolution, multimodal data. Why, then, are these methods not yet commonplace, and why do many scientists struggle to harness them? We believe the fundamental obstacle lies less in the choice of ML model and more in their implementation, dissemination, and overall \textit{utility} to the materials science community. By addressing these issues through science-driven engineering, we can develop practical tools to integrate autonomy into routine experimentation.

Nearly a decade ago, our group began to consider the barriers to autonomy in electron microscopy, and materials science more broadly. We have identified these barriers as falling into two main categories: cultural and practical.\cite{Spurgeon.10.1038/s41563-020-00833-z} Culturally, many scientists are justifiably wary of ``black box'' models that operate via opaque logic. Most ML models are not directly interpretable (at least in a traditional sense), and it is difficult to determine whether a given experiment might encounter an out-of-distribution scenario where such models would fail. Practically, many scientists lack formal training in data science or its lingua franca, Python. Consequently, many scientists struggle to implement these models and select key parameters that can significantly impact their performance. Just as importantly, most microscope hardware has not been designed with autonomy in mind, aside from specific, narrow, and vendor-defined use cases. These combined factors have hindered the widespread adoption of AI/ML in microscopy.

We have spent considerable time addressing autonomous science from both the cultural and practical angles. We have identified four critical criteria for the community to embrace autonomy: First, AI/ML models must be understandable at some level, fostering trust in their operation. Second, they must be powerful, offering a synthesis or interpretation of data that surpasses human capabilities. Third, they must be usable during an experiment, allowing for real-time adjustments and immediate feedback. Fourth, they must be deployable within an autonomous system integrated into the microscope hardware. By fulfilling these criteria, we have engineered practical autonomous microscopy systems that are transforming how materials studies are conducted.

This section describes a progression of autonomy analogous to the ``co-pilot'' and ``full self-driving'' concepts used in autonomous vehicles. We first introduce a co-pilot for ``nanocartography''---the art of navigating samples within the microscope with respect to the stage, linking the physical world to reciprocal space. While this skill is traditionally acquired through apprenticeship, we have devoted considerable effort to both codifying this process and developing intuitive interfaces to facilitate it. Building upon this co-pilot concept, we then describe the design of a fully autonomous microscope system. We detail how ML can inform real-time decision-making, how feedback is implemented, and which aspects of standard microscopy protocols are most amenable to autonomous operation. Taken together, these approaches underscore the need not only for high-performance models, but also for those that can be practically implemented in meaningful ways.

\subsection{From Crystallographic Co-Pilot\ldots}\label{sec:nanocarto}

One of the first things a microscopist needs to learn is how to orient themselves in the often complex nanoworld. While the first assumption about TEM samples is that they are nearly infinitely thin, in reality even samples at 50~nm thick can contain multiple phases and interfaces that require careful inteprretation. However, traditional pedagogy relies on observational learning, leading to a wide variety of practices that are difficult to propagate precisely. Moreover, nearly every operation, from sample navigation to tilting and focusing, becomes more challenging at high magnification.\cite{Fiedler.10.1093/micmic/ozad108} When combined with beam-sensitive or complex samples whose projections are not easily interpretable, it becomes clear that a strong grasp of nanocartography is a prerequisite for any realistic automation. Precise, or as practically achievable, control of the stage is an absolute necessity. 

Building upon prior efforts in digital notebooks and practical microscopy tools, we designed the NanoCartographer package to address several key limitations.\cite{Olszta.10.69761/dnka1581} First, it is crucial to record orientation information relative to the sample's frame of reference. For example, in grain boundary microstructures, understanding the sample's loading orientation within the instrument influences how we subsequently tilt and align within a specific grain. This information is frequently lost, leading to wasted time and effort. Furthermore, this information is not easily conveyed to collaborators, hindering the implementation of federated microscopy experiments. Second, we need the ability to readily translate between the instrument's coordinate system and a relevant crystallographic system. As any materials science student knows, calculating crystal planes and vectors is straightforward for cubic systems but becomes considerably more difficult for other crystal types, such as hexagonal or triclinic systems. We must become fluent in the transition between instrument and crystal coordinate systems. Finally, our ability to access crystallographic zone axes is constrained by the ability to translate the stage while at a tilt; given that the microscope pole piece gap is only 5--10 mm, we are often limited to a few hundred microns of travel at zero tilt, which can be reduced to mere microns at high tilt angles. Thus, it is critical to know the available range of motion at any given time.

\begin{figure}[h!]
    \centering
    \includegraphics[width=0.9\textwidth]{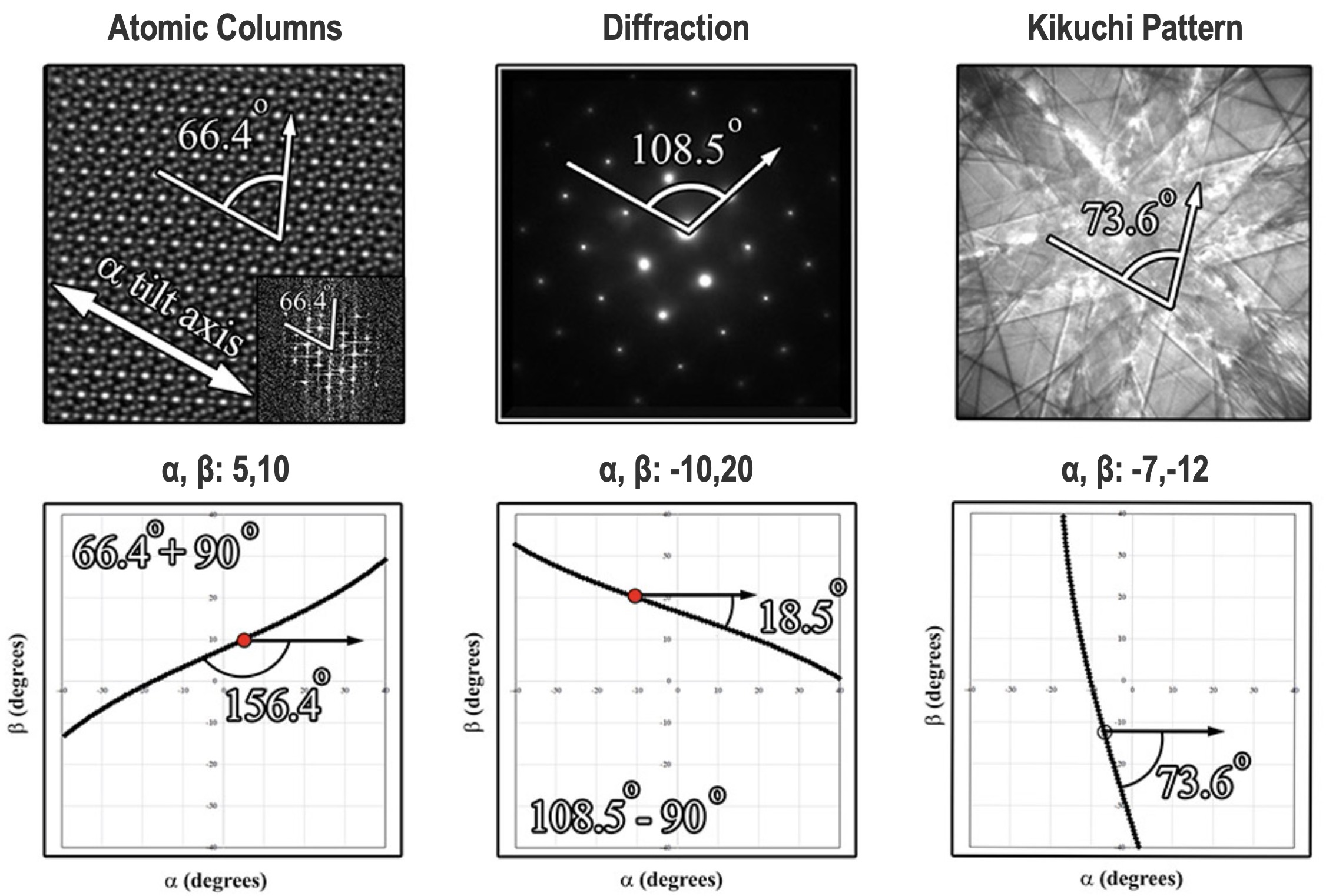}
    \caption{\textbf{A crystal to microscope interpreter abstracts instrument control.} NanoCartographer aids in the translation from instrument to meaningful crystallographic information for imaging and diffraction data, improving the speed and reproducibility of experiments. Adapted from Olszta et al. under CC-BY 4.0 license.\cite{Olszta.10.69761/dnka1581}}
    \label{fig:copilot}
\end{figure}

As shown in Fig. \ref{fig:copilot}, NanoCartographer provides the microscopist with the ability to develop a practical map of their sample. The software comprises a digital notebook for recording raw, instrument-level orientation information, which is then translated into a vector representation of the crystal system. The program also offers rapid analytical tools, including precise crystal tilting, unknown crystal generation, accurate tilting of non-crystallographic features, predictive grain boundary analysis, and crystal/interface calculations. In practice, this facilitates challenging experiments, such as oblique tilting around specific grain boundaries and interfaces. The digital notebook enables easy, reliable sharing of orientation information between collaborators, in alignment with FAIR (Findable, Accessible, Interoperable, and Reusable) data principles.\cite{Wilkinson.10.1038/sdata.2016.18}

In computing parlance, NanoCartographer acts as an interpreter, translating semantically meaningful, high-level materials queries into low-level instrument coordinates and vice-versa. The simplicity of the interface belies the complex vector mathematics relating the sample, instrument, and crystallographic frames of reference. This abstraction of low-level instrument control is one of the most important, yet under-appreciated, steps in the transition to full autonomy. By creating a useful crystallographic co-pilot, we can bridge the gap between human- and machine-based workflows.

\subsection{\ldots To Full Self-Driving Microscope}\label{sec:autoem}

The previously described microscopy co-pilot is transforming current microscopy practices. But what does the future of microscopy hold? We must first recognize that microscopes have been designed around a human-in-the-loop paradigm for nearly a century. Consequently, every aspect of the instrument, from command input to stage operation and data display, is human-centric. Embedding autonomy into such a system necessitates a fundamental redesign of control and reasoning. A decade ago, this would have been nearly impossible; however, in recent years, instrument vendors have begun to provide access to hardware through new application programming interfaces (APIs).\cite{Olszta.10.1017/s1431927622012065, Potocek.10.1002/aisy.202300745} We have collaborated closely with these vendors, integrating ML-based reasoning with sparse data to achieve practical control of the microscope. The outcome is a platform that emulates, albeit imperfectly, the cognitive processes of a human scientist.
    
\begin{figure}[H]
    \centering
    \includegraphics[width=0.44\textheight]{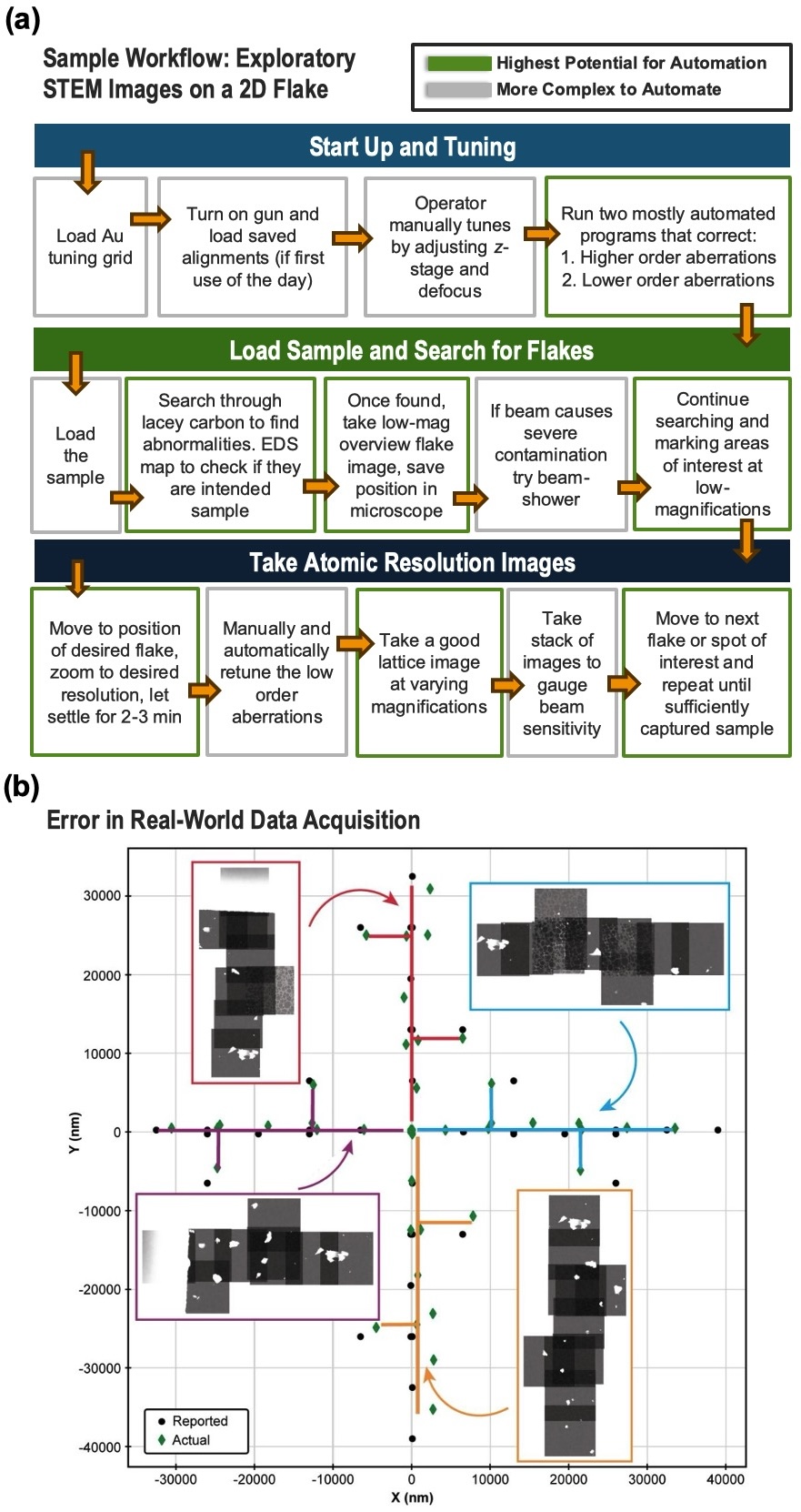}
    \caption{\textbf{Highlighting the gap between idealized autonomous workflows and their practical implementation.} (a) Step-by-step logic for a prototype experiment in analyzing 2D materials flakes, with areas for AI improvement indicated. (b) Results of automated stage movement implemented using the AutoEM system, revealing errors in positioning and challenges in post-processing. Adapted from Fiedler et al. under CC-BY 4.0 license.\cite{Fiedler.10.1093/micmic/ozad108}}
    \label{fig:fsd}
\end{figure}

The first step to automating a microscopy workflow is observing and capturing the current process. Acquiring a single high-resolution STEM image is a complex multi-step process that involves many decisions traditionally guided by the intuition of an experienced microscopist. To translate the non-standardized work of a researcher into a program, each decision from microscope initialization onward must be systematically documented. While experiments vary based on desired data and sample type, many workflows share common steps. Fig. \ref{fig:fsd} shows how to distill the STEM imaging of 2D materials such as MXenes into actionable logic. Fine-tuning of beam aberrations, selecting relevant sample regions, assessing the image quality, ensuring sufficient data collection, and managing contamination all require careful consideration and precise adjustments.

We have found that the best practice is to automate this workflow in stages. When determining which steps to automate first, we must consider several factors including the desired level of automation, number of real-time decisions, and API limitations, which differ by vendor and instrument. After the initial set up, it is common to search through a sample and identify regions of interest. To do this, we require an ML algorithm running in parallel with the microscope to be able to identify such areas. In principle, programming a montage-like stage movement for this sample sweep is straightforward, as is acquiring an image of the area and saving the position for easy rediscovery. Most APIs now allow for easy adjustment of optical parameters and automated image acquisition with the ability to control the number of images, size, and dwell time. Automating stage movement, image acquisition, and predefined parameter adjustments is relatively straightforward. However, acquiring high-quality data using such APIs still presents a major challenge. 

\begin{figure}[h]
    \centering
    \includegraphics[width=\textwidth]{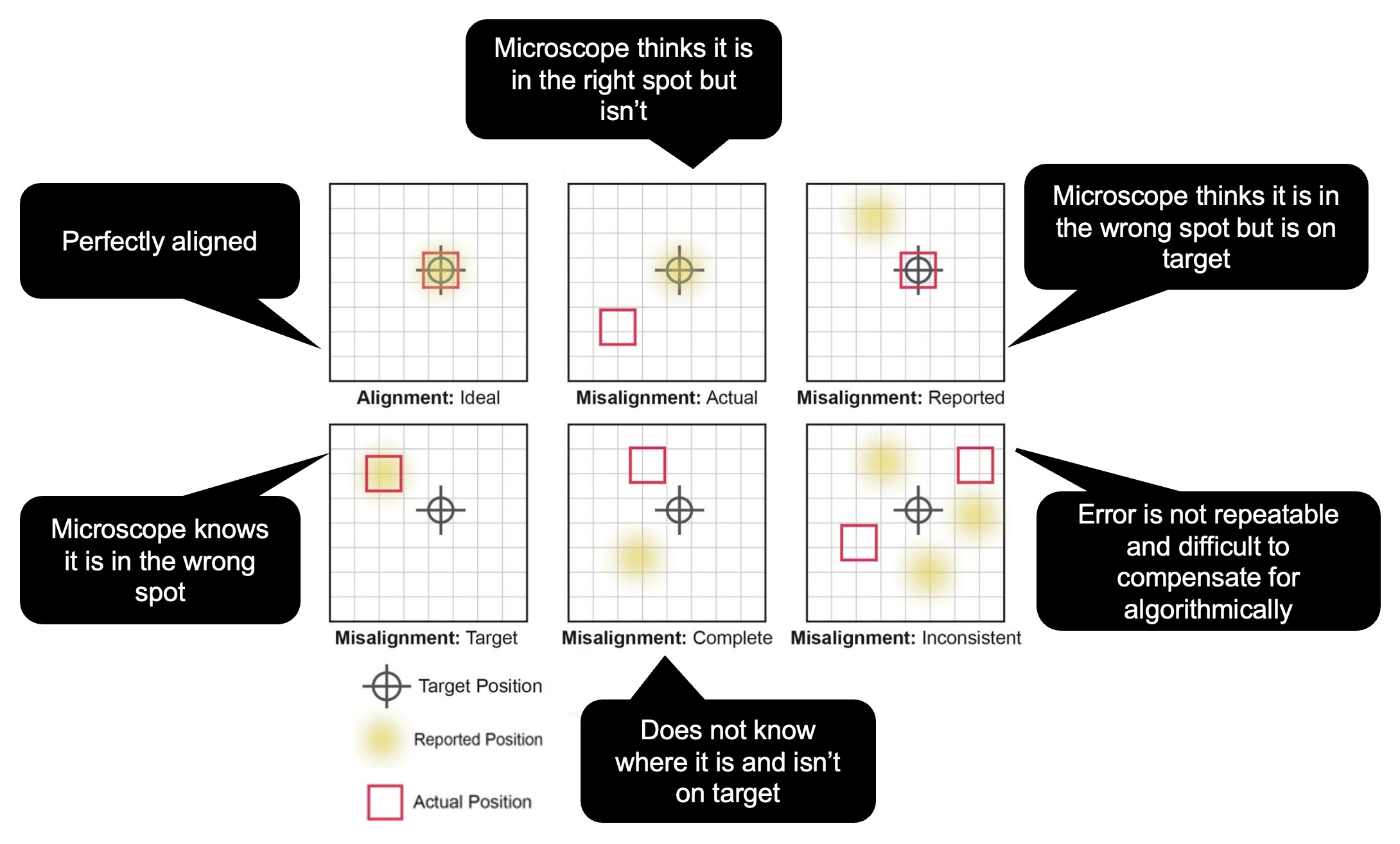}
    \caption{\textbf{Understanding the challenges of real-world control.} Executing desired workflows is complicated by imprecise and hysteretic instrument behavior, which make it difficult to program autonomous experiments. Adapted from Fiedler et al. under CC-BY 4.0 license.\cite{Fiedler.10.1093/micmic/ozad108}}
    \label{fig:AlignmentReworked}
\end{figure}

Both hardware and software limitations represent major barriers to fully automated microscopy.\cite{Fiedler.10.1093/micmic/ozad108} A key challenge is hysteresis and instability of mechanical stage movement. We have found that programmed movements do not always result in precise or accurate physical stage responses. This error compounds for chains of movement which is unacceptable when working at the atomic scale. Fig. \ref{fig:fsd} shows automated montages that were supposed to be taken at equidistant steps along $x$ and $y$ stage directions. The actual data displays substantial variance in the step sizes along with the impurity of the single axis movement. The possible discrepancies between commanded, reported and actual stage movements are displayed in Fig. \ref{fig:AlignmentReworked}. The appropriate corrective approach depends on the nature of the misalignment. Ideally the mechanical components would be upgraded for greater precision, but until then the only option is integrating software-based error correction.

Software development for automation varies significantly in maturity, depending on the instrument and specific task. While both first- and third-party vendors offer automated correction for low- and higher-order aberrations, as well as drift correction, these functionalities are often limited in functionality. Similarly, existing auto-focusing functions are typically designed for a narrow range of settings and may not be suitable for many experimental conditions. Furthermore, the proprietary nature of these tools hinders their seamless integration. Combining the capabilities of these disparate tools would significantly enhance our ability to automate a wider range of experiments with greater efficiency and accuracy. However, the reality today is that researchers must develop bespoke solutions for even relatively routine tasks, a situation that has hampered progress and created barriers to wider adoption of automation in microscopy. To accelerate the advancement of autonomous microscopy, a concerted effort is needed to develop open, modular, and interoperable software solutions that empower researchers to readily adapt and combine automation functionalities, fostering a more collaborative and innovative ecosystem.

Further acceleration in these automation capabilities can be addressed through development of virtual microscope environments or digital twins. Iterative testing of software workflows on physical microscope systems can be time-consuming and blocks the instrument from being used in current experimental campaigns. Simulated environments provide a controlled platform for first stage evaluation and refinement of algorithms; this is particularly important for more complicated AI-driven workflows or policies developed via reinforcement learning (RL) that require extensive evaluation in-the-loop. For example, such simulators have been used for training RL agents to align the bright field disk on a detector\cite{Xu.10.1017/s1431927622012193} and in Lorentz imaging of skyrmions.\cite{McCray.10.1063/5.0197138}  However, the realization of such virtual environments, and their effective application, critically depends on the parallel development of FAIR software and hardware platforms, such that researchers can construct digital twins that accurately reflect real-world microscopy setups.

\section{Conclusions}

This perspective has outlined the trajectory of our group's work in autonomous microscopy and our vision for its implementation. However, significant opportunities remain for developing and implementing new solutions.

First, models must continue to be designed with the specific challenges of microscopy and materials science in mind. Initial work has highlighted both the promise and the challenges associated with implementing multimodal models, most notably the lack of suitable encoders, benchmark datasets, effective data fusion methods, and approaches that account for the cost of data acquisition. The development of new models will offer improved discovery power, greater interpretability, and ultimately, enhanced confidence in describing materials systems.

Second, we lack community-wide standard benchmark datasets for training and validation. While microscopy will likely always be plagued by the projection problem (described in Section \ref{fig:precision}), it should nevertheless be possible to curate benchmark datasets for key scenarios such as 2D materials imaging, polycrystalline metals, thin films, etc. With such datasets, we can design new encoders and rigorously validate model performance, as is common practice in many other areas of data science. Here, autonomy will eventually assist in collecting and curating data in a standardized format, but a consensus is still needed on how best to structure and disseminate this data.

Third, we must change the mentality around hardware design. For years, the community has prioritized implementing aberration correctors, brighter sources, and faster cameras, while overlooking stage reproducibility and software workflows for autonomous data collection. We should adopt practices from industry regarding standardization of analyses, demand open access to instrument control and software, and structure our experiments around machine- rather than human-in-the-loop paradigms. In so doing, we will collectively incentivize instrument manufacturers to make the required changes.

What is the future for autonomy in microscopy and materials science in general? One can envision a day when truly standardizable, reproducible campaigns of materials analysis are conducted, using models that have been tuned to the problem of detecting defects, measuring phase distributions, or understanding phase transitions on the basis of time-varying and multimodal data. This information will enable us to automatically derive more complete and accurate physical models, providing actionable insights that can guide materials synthesis and processing. Indeed, we are already witnessing this transition, as evidenced by the numerous tools and approaches that are bridging the gap towards true autonomy. When---and if---that goal is fully realized, the role of the scientist will evolve to orchestrate a network of autonomous agents capable of proposing and testing hypotheses at a scale far exceeding our current capabilities. The result will be a deeper understanding and mastery of the complex atomic world that will fuel innovation and scientific discovery.

\clearpage

\section{Acknowledgements}

This work was authored in part by the National Renewable Energy Laboratory (NREL), operated by Alliance for Sustainable Energy, LLC, for the U.S. Department of Energy (DOE) under Contract No. DE-AC36-08GO28308. G.G., A.S., M.S., A.G., H.E., and S.R.S. were supported by the Operando to Operation (O2O) Laboratory Directed Research and Development (LDRD) program at NREL. M.J.O. and K.R.F. were supported by the Adaptive Tunability for Synthesis and Control via Autonomous Learning on Edge (ATSCALE) LDRD at Pacific Northwest National Laboratory (PNNL). PNNL is a multiprogram national laboratory operated for the U.S. Department of Energy (DOE) by Battelle Memorial Institute under Contract No. DE-AC05-76RL0-1830. The views expressed in the presentation do not necessarily represent the views of the DOE or the U.S. Government. The U.S. Government retains and the publisher, by accepting the article for publication, acknowledges that the U.S. Government retains a nonexclusive, paid-up, irrevocable, worldwide license to publish or reproduce the published form of this work, or allow others to do so, for U.S. Government purposes. B.C.W and B.A. acknowledge funding support from the U.S. NSF, award number CMMI-2134607.

\section{Author Contribution Statement}

G.G. led the writing of the section on atomic-resolution analysis of MXenes and contributed to the organization of the review. A.S. led the writing of the section detailing protocols for high-resolution workflows and editing of the figures. M.S. collected and analyzed the MXene STEM data. A.G. and H.E. contributed to ML development for MXene analysis. B.C.W. and B.A. synthesized the MXene samples. M.J.O. and K.R.F. led the writing of the section on nanocartography. S.R.S. oversaw the project, wrote the sections on AutoEM and multimodal data integration, and led the conceptualization of the review. All authors contributed to the editing and refinement of the manuscript.

\section{Data Availability Statement}

The Ti$_3$C$_2$ MXene for this work was synthesized using previously published methods\cite{Thakur.10.1002/smtd.202300030} and the data on MXene analysis are available from the authors upon reasonable request. All other data are available in published repositories and manuscripts.

\section{Conflict of Interest Statement}

The authors have no conflicts to disclose.

\clearpage 

\section{Highlight Image}

\begin{figure}[h!]
    \centering
    \includegraphics[width=\textwidth]{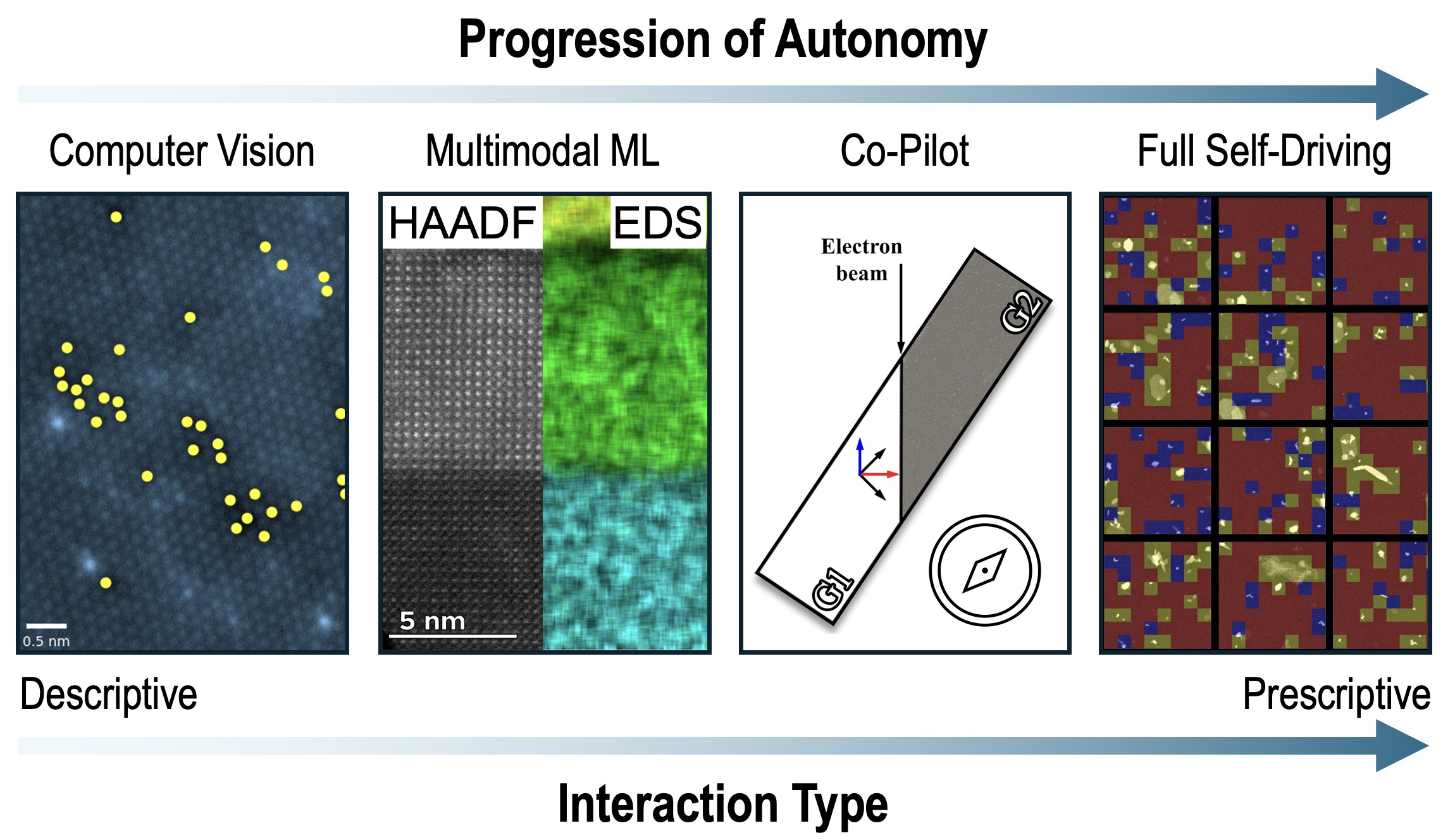}
    \label{fig:toc}
\end{figure}

\clearpage

\bibliography{references}
\bibliographystyle{unsrturl}

\end{document}